\documentclass[journal = aaemcq, manuscript=article, layout=twocolumn]{achemso}

\usepackage{amsmath}
\usepackage{graphicx}
\usepackage{dcolumn}
\usepackage[english]{babel}

\usepackage{color}
\usepackage{float}

\usepackage{amssymb}

\let\oldmaketitle\maketitle
\let\maketitle\relax

\author{Sergei M. Butorin}
\email{sergei.butorin@physics.uu.se}
\affiliation{Condensed Matter Physics of Energy Materials, X-ray Photon Science, Department of Physics and Astronomy, Uppsala University, P.O. Box 516, SE-751 20 Uppsala, Sweden}

\title{On the band gap variation in CH$_3$NH$_3$Pb(I$_{1-x}$Br$_x$)$_3$}


\begin{document}


\twocolumn[
\begin{@twocolumnfalse}
\oldmaketitle

\begin{abstract}
The electronic structure and the band gap behavior of CH$_3$NH$_3$Pb(I$_{1-x}$Br$_x$)$_3$ for $x$=0.25, 0.33, 0.50, 0.67, 0.75, 1.00 were studied using the full-relativistic density-functional-theory calculations. A combination of the parameter-free Armiento-K\"{u}mmel generalized gradient approximation exchange functional with the nonseparable gradient approximation Minnesota correlation functional was employed. The calculated band gap sizes for the CH$_3$NH$_3$Pb(I$_{1-x}$Br$_x$)$_3$ series were found to be similar to the experimentally measured values.  While the change of the optimized lattice parameter with an increasing Br content can be described by a linear fit, the calculated band gap variation exhibits rather a quadratic-like behavior over the $x$ region of the cubic crystal structure. While the experimental reports are divided on whether the bowing parameter value is being very small or significant, our calculated results support the latter case.

\end{abstract}

\end{@twocolumnfalse}
]

\begin{tocentry}
\includegraphics[height=4.5 cm]{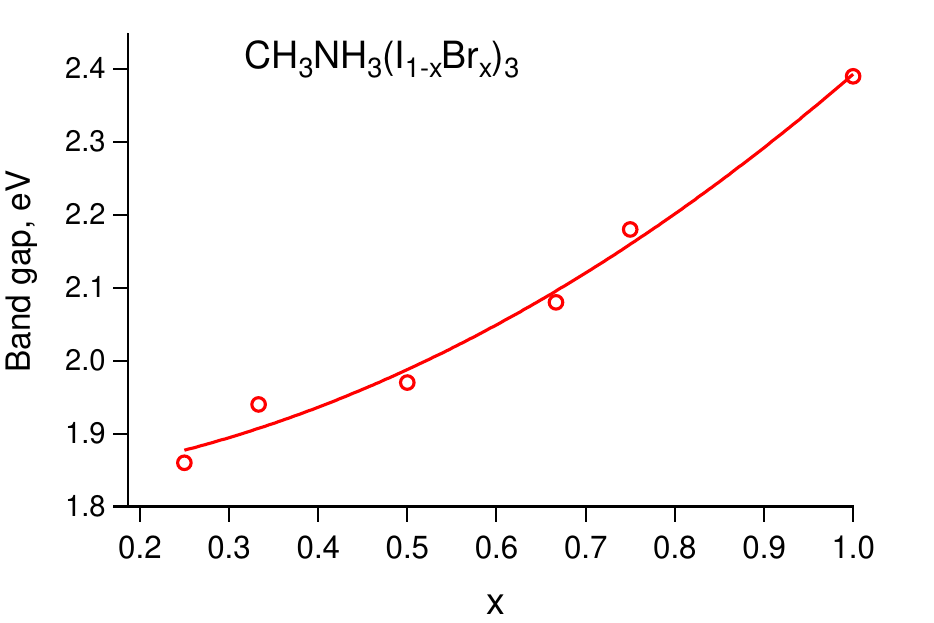}
\end{tocentry}

\section{Introduction}

Hybrid metal halide perovskites (HaPs) are considered to be advanced materials for solar cell applications \cite{Kojima}. One of the valuable properties of HaPs is an ease of the adjustment of their band gap by an interchange of the halogen atoms. In particular, in the mixed halide MAPb(I$_{1-x}$Br$_x$)$_3$ system (MA stands for methylammonium), the band gap varies from $\sim$1.5 eV for $x$=0 to $\sim$2.3 eV for $x$=1. Although, that was demonstrated in a number of publications \cite{Noh,Kulkarni,Fedeli,Sadhanala,Park,Atourki,Misra,Zhang,Wang,Nakamura}, the published data are not always consistent with each other. For example, different $x$ values were reported for the observed phase transition in the MAPb(I$_{1-x}$Br$_x$)$_3$ series from the MAPbI$_3$-like tetragonal structure to the cubic one. The dependence of the increasing band gap on the Br content, replacing I, was found to be linear \cite{Park,Zhang} or quadratic-like \cite{Noh,Fedeli,Atourki,Wang,Nakamura} with a positive bowing parameter \cite{Hill} value (coefficient for the $x^2$ term) in the fitting equation. Moreover, Zhang \textit{et al.} \cite{Zhang} reported a fit with a negative bowing parameter. It was also hinted that the non-linear behavior of the band gap with increasing Br fraction may be connected to a non-equivalent  replacement of I with Br or to a deviation from the intended Br quantity during the sample synthesis as suggested by alternative methods of the estimation of the Br content.

A number of the density functional theory (DFT) calculations were also performed \cite{Mosconi,Chen,Chang,Martynow,Roy} to study the crystal and electronic structure of MAPb(I$_{1-x}$Br$_x$)$_3$. However, in most cases, the spin-orbit coupling (SOC) was not included in the DFT analysis. SOC is strong in the Pb-based HaPs and significantly affects the density of states (DOS) close to the conduction band minimum (CBM) in these materials. Therefore, it is important to take into account SOC for HaPs. Chang \textit{et al.} \cite{Chang} applied SOC as well as hybrid functional HSE06 in their calculations but investigated only MAPbI$_2$Br and MAPbIBr$_2$ compounds out of the whole MAPb(I$_{1-x}$Br$_x$)$_3$ series. Note that most of the cited computational work was based on utilizing the tetragonal-like unit cells while recent experiments \cite{Wang,Nakamura} for single crystals and polycrystalline samples indicated the cubic structure of MAPb(I$_{1-x}$Br$_x$)$_3$ for $x$ above 0.2.

The intension of the present work was to investigate the electronic structure and the band gap changes in the MAPb(I$_{1-x}$Br$_x$)$_3$ series using the DFT method taking into account SOC. The recently suggested approach \cite{Butorin} combining the the parameter-free Armiento-K\"{u}mmel generalized gradient approximation (AK13-GGA) exchange functional \cite{Armiento} with the nonseparable gradient approximation Minnesota correlation functional (GAM) \cite{Yu} was applied. This allows for an efficient band gap estimation with accuracy similar to the GW approximation method but at the computational cost of conventional DFT. The described approach creates an opportunity for the effective assessment of the electronic structure of doped HaPs when the use of supercells for the DFT calculations is necessary.

\section{Results and discussion}

\begin{figure*}
\includegraphics[width=\textwidth]{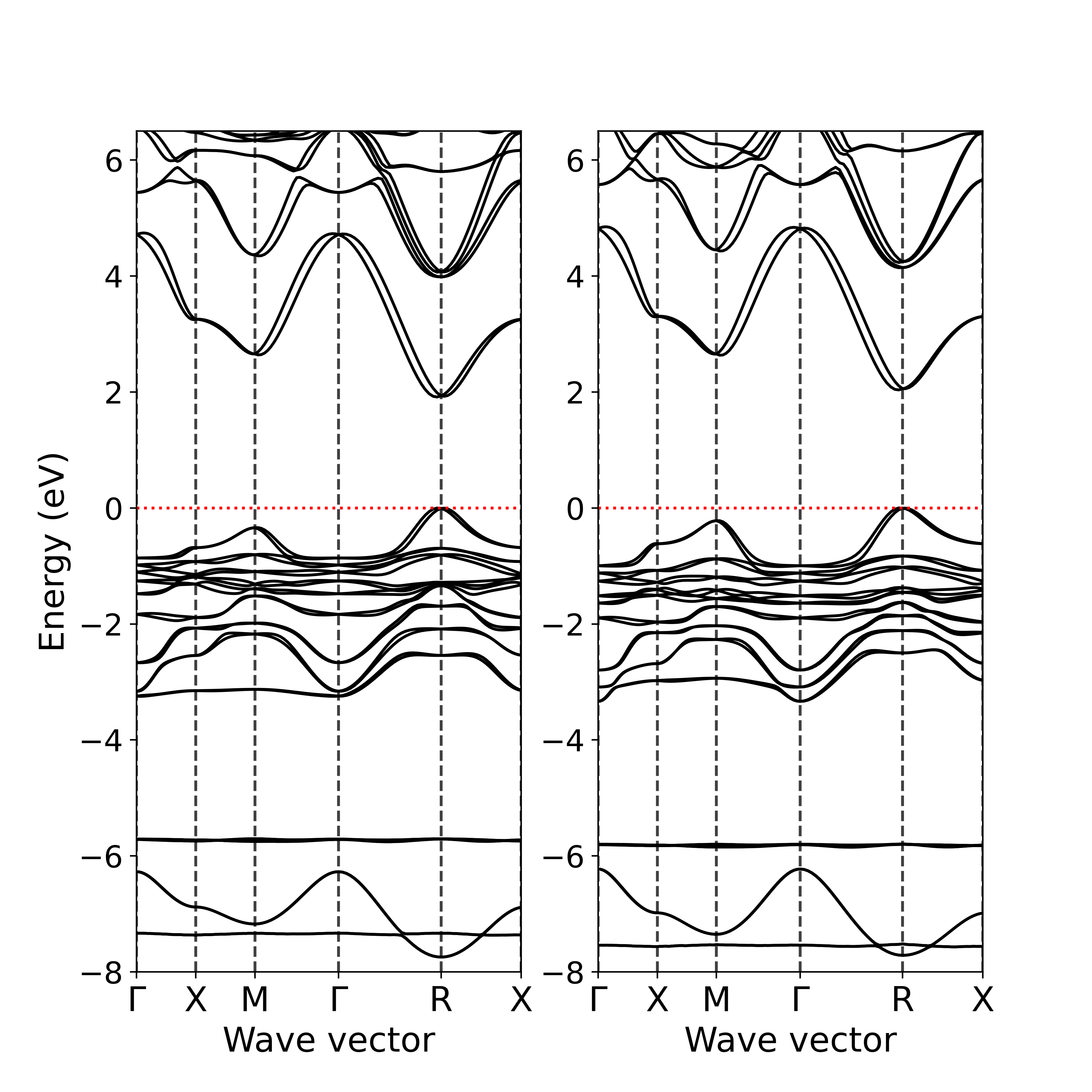}
\caption{Band structure of pseudocubic MAPbI$_2$Br (left hand side) and MAPbIBr$_2$ (right hand side) calculated at the level of AK13/GAM theory. Zero eV is at the valence band maximum. \label{MAPbIBr_bands}}
\end{figure*}

\begin{table*}
\centering
\caption{AK13/GAM calculated and experimental band gaps of MAPb(I$_{1-x}$Br$_x$)$_3$ (in units of eV).}
\begin{tabular}{cccc}
Br content, $x$&Calculation&Experiment (Ref.\cite{Nakamura})&Experiment (Ref.\cite{Wang})\\
\hline
0.00&1.42 (Ref.\cite{Butorin})&1.65&1.56\\
0.25&1.86&1.83&1.71\\
0.33&1.94&1.87&1.77\\
0.50&1.97&1.97&1.88\\
0.67&2.08&2.09&2.00\\
0.75&2.18&2.15&2.05\\
1.00&2.39 (Ref.\cite{Butorin})&2.37&2.23\\
\end{tabular}
\label{table1}
\end{table*}

The electronic structure of MAPb(I$_{1-x}$Br$_x$)$_3$ was studied for $x$=0.25, 0.33, 0.50, 0.67, 0.75, 1.00 compositions. The $x=$0.33 and 0.67 systems correspond to MAPbI$_2$Br and MAPbIBr$_2$ compounds, respectively. For this purpose, the 2x2x2 supercells (96 atoms in total) were constructed where the corresponding number of Br and I atoms interchanged to obtained the required composition. In the case of MAPbI$_2$Br and MAPbIBr$_2$, the conventional unit cells were used in the DFT calculations for simplicity. Since the most recent experimental reports \cite{Wang,Nakamura} on the crystal structure of MAPb(I$_{1-x}$Br$_x$)$_3$ indicate that the tetragonal to cubic transition takes place for $x$$\geq$0.2, in present work, the DFT calculations of the electronic structure were performed for optimized pseudocubic crystal structures for all described compositions. In other words, the cubic crystal structure region of the MAPb(I$_{1-x}$Br$_x$)$_3$ phase diagram was sampled.

\begin{figure*}
\includegraphics[width=\textwidth]{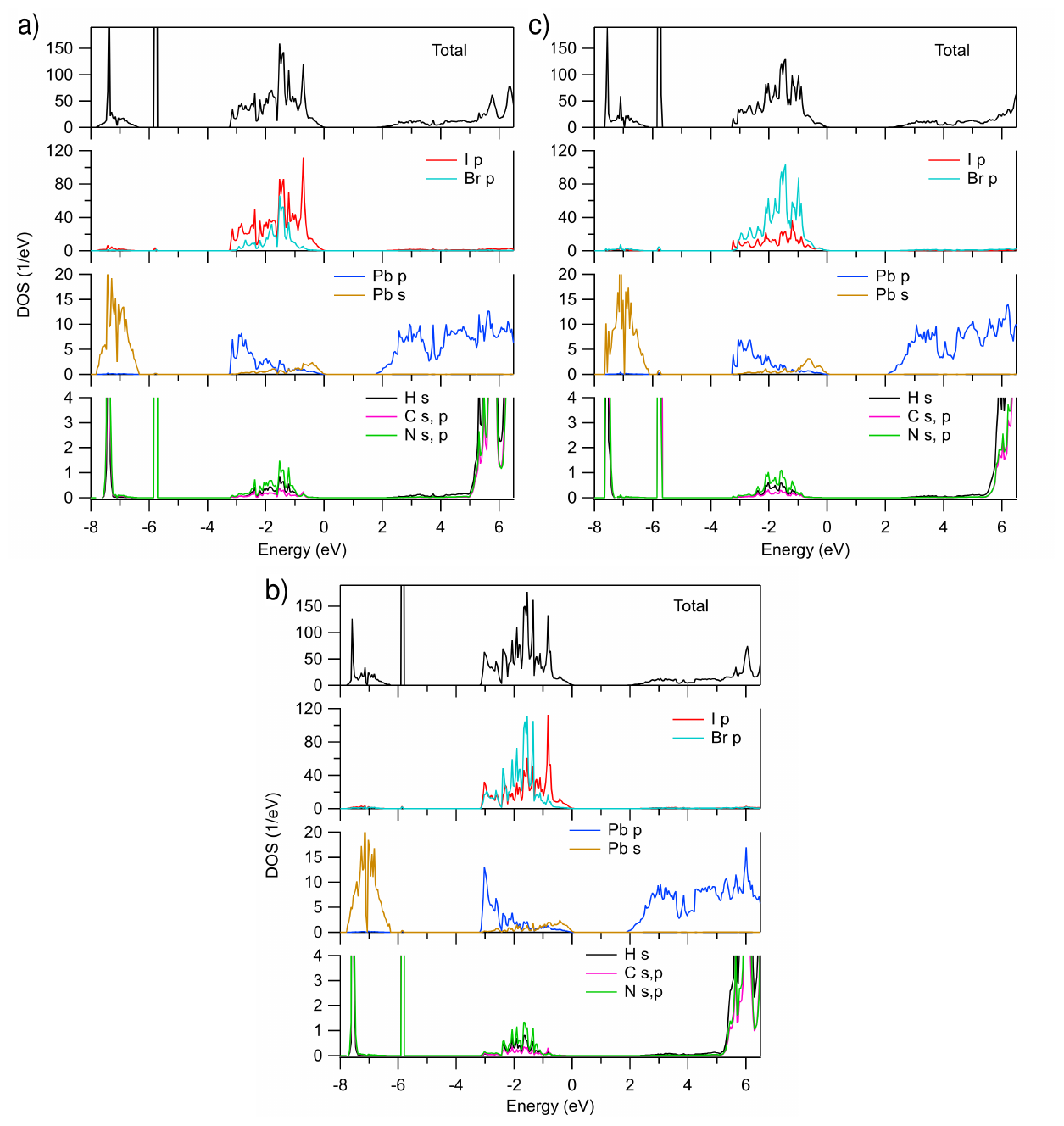}
\caption{Total and partial densities of states of pseudocubic MAPb(I$_{1-x}$Br$_x$)$_3$ calculated for a) $x$=0.25, b) $x$=0.50, and c) $x$=0.75, respectively. Zero eV is at the valence band maximum. \label{DOS_MAPbBrI}}
\end{figure*}

Figure~\ref{MAPbIBr_bands} displays the calculated band structure for MAPbI$_2$Br and MAPbIBr$_2$. It is evident that having both smaller and larger fractions of I or Br atoms in the cubic crystal structure does not lead to any dramatic changes in the band structure and band dispersion. Besides an increasing direct band gap at high-symmetry point R from 1.94 to 2.08 eV upon going from MAPbI$_2$Br to MAPbIBr$_2$, the topmost occupied band at high-symmetry point M appears to be closer to valence band maximum (VBM) in MAPbIBr$_2$ than in MAPbI$_2$Br. Some differences are also observed for the unoccupied bands above 6 eV, in particular at high-symmetry points X, M and $\Gamma$ and for the X$-$M$-$$\Gamma$ path. These calculated band structures are in good correspondence with that published earlier \cite{Butorin} for cubic MAPbBr$_3$ which was also calculated using the AK13/GAM combination.

Table~\ref{table1} compares the calculated band gap sizes of MAPb(I$_{1-x}$Br$_x$)$_3$ for various $x$ with experimental ones \cite{Wang,Nakamura} for single crystals and polycrystalline samples. Besides the optically measured actual data, the experimental band gap sizes for some compositions were derived from equations used to fit the band gap dependence on $x$ in Refs.\cite{Wang,Nakamura}. The values for tetragonal MAPbI$_3$ are also included in Table~\ref{table1} to span the full $x$ range. Note that the AK13/GAM calculated band gap for the 2x2x2 supercell of MAPbBr$_3$ was the same as that in Ref.\cite{Butorin}.

An inspection of Table~\ref{table1} reveals that calculated band gaps are in agreement with experimental results by Nakamura \textit{et al.} \cite{Nakamura} but appear to be somewhat overestimated when compared with data published by Wang \textit{et al.} \cite{Wang}. Overall, there are also some variations in the band gap sizes among experimental reports \cite{Noh,Kulkarni,Fedeli,Sadhanala,Park,Atourki,Misra,Zhang,Wang,Nakamura} when the same compositions are compared. That may be connected to the varying quality of the samples and different methods of their synthesis.

The AK13/GAM calculated total and partial densities of states (DOS) of MAPb(I$_{1-x}$Br$_x$)$_3$  for $x$=0.25, 0.50, and 0.75 are shown in Figure~\ref{DOS_MAPbBrI}. As expected, the valence band is dominated by I and Br $p$ states. However, their distribution over the extent of the valence band differs. The gravity center of the Br $p$ DOS appears to be at lower energies and the I $p$ states tend to contribute more in vicinity of VBM, at least for $x\leq0.5$. Such a behavior can be one of the reasons why the band gap in MAPbI$_3$ is smaller than in MAPbBr$_3$. The calculated results are in line with observed changes in x-ray photoelectron spectroscopy (XPS) of the valence band of MAPbI$_3$ doped with Br (Ref. \cite{Park}). Upon increasing Br content, the spectra shift to the higher binding energies and change their shape so that the spectral weight decreases on the low binding energy side. In turn, the calculated Pb $p$ and $s$ DOSs in the valence band qualitatively similar to those in MAPbI$_3$ and MAPbBr$_3$ where the Pb $p$ and $s$ states mainly contribute at the bottom and top of the valence band, respectively. The fractions of the H $s$, C and N $s$, $p$ states in the valence band are quite small and negligible at VBM.

\begin{figure}
\includegraphics[width=\columnwidth]{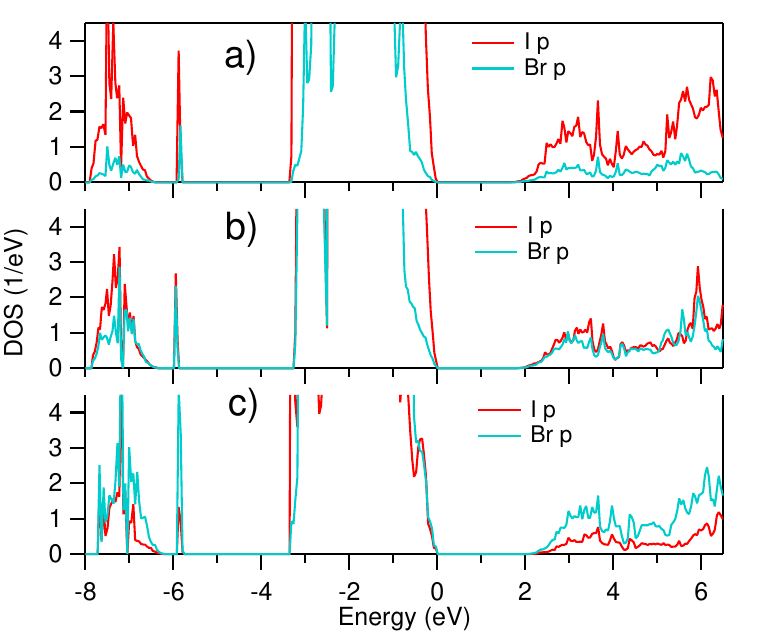}
\caption{Zoom onto unoccupied I $p$ and Br $p$ densities of states of pseudocubic MAPb(I$_{1-x}$Br$_x$)$_3$ calculated for a) $x$=0.25, b) $x$=0.50, and c) $x$=0.75, respectively. \label{DOS_Br_I}}
\end{figure}

Since SOC affects strongly the unoccupied states in vicinity of CBM, a zoomed view on the AK13/GAM calculated unoccupied I and Br $p$ DOSs of MAPb(I$_{1-x}$Br$_x$)$_3$ is depicted in Figure~\ref{DOS_Br_I}. A comparison of results in Figures~\ref{DOS_MAPbBrI} and \ref{DOS_Br_I} shows that the shape of the unoccupied Pb $p$ DOS is closely repeated by the distribution of the unoccupied I and Br $p$ DOSs, despite the contributions by both I, Br $p_{\pi}$ and $p_{\sigma}$ states (Ref.\cite{Man}) which are expected to have a different degree of hybridization.

\begin{figure}
\includegraphics[width=\columnwidth]{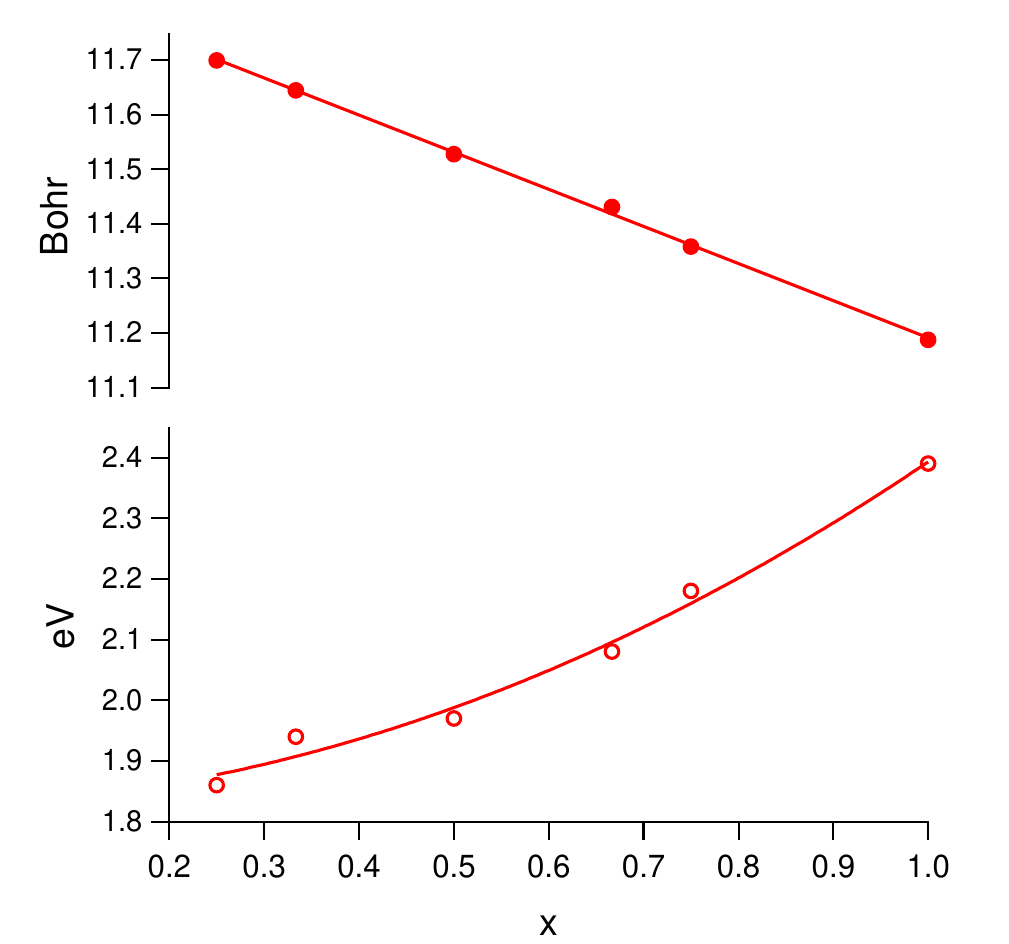}
\caption{The optimized lattice parameter (solid markers) and band gap size (open markers) of MAPb(I$_{1-x}$Br$_x$)$_3$ in the cubic structure region. \label{Lattice}}
\end{figure}

Figure~\ref{Lattice} displays the dependence of the optimized lattice parameter and band gap size of MAPb(I$_{1-x}$Br$_x$)$_3$ on $x$ in the cubic structure region. While the change of the lattice parameter value with an increasing Br content can be described by a linear fit, the band gap ($E_g$) variation exhibits rather a quadratic-like behavior. The result of the least-squares fit can be described by the following equation:

\begin{eqnarray}
E_g = 0.490x^2+0.074x+1.828.
\end{eqnarray}

\noindent The calculated band gap size for $x=0.33$ is only noticeably off the fitting curve in the latter case. While the experimental reports are divided on whether the bowing parameter value is being very small \cite{Fedeli,Atourki,Wang} or significant \cite{Noh,Nakamura}, our calculated results (bowing parameter value 0.490) support the latter case.

\section{Methods}

The DFT calculations were performed in the full-relativistic mode using the Quantum Espresso v.6.8 code \cite{Giannozzi}. A combination of AK13/GAM (as they are defined in the LibXC v.5.1.6 library \cite{Lehtola}) was applied for MAPb(I$_{1-x}$Br$_x$)$_3$ with varying $x$. The full-relativistic norm-conserving PBE (Perdew, Burke, and Ernzerhof \cite{Perdew}) pseudopotentials for hydrogen, carbon, nitrogen, bromine, iodine and lead were generated by the code of the ONCVPSP v.4.0.1 package \cite{Hamann} using input files from the SPMS database \cite{Shojaei}. The pseudopotentials were generated without non-linear core correction. An additional feature of this ONCVPSP version is its ability to check for positive ghost states. The valence configurations for the pseudopotentials were defined as 1s$^1$ for H, 2s$^2$2p$^2$ for C, 2s$^2$2p$^3$ for N, 4s$^2$4p$^5$ for Br, 5s$^2$5p$^5$ for I, and 5d$^{10}$6s$^2$6p$^2$ for Pb. The plane-wave cut-off energy was set to 60 Ry. The convergence threshold for density was 1.0x10$^{-12}$ Ry. The Van der Waals correction was applied using Grimme's D2 method \cite{Grimme}. The Brillouin zone was sampled using the Monkhorst-Pack scheme \cite{Monkhorst} and sizes of the shifted $k$-point mesh were chosen to be 4x4x4 for the 2x2x2 supercells of MAPb(I$_{1-x}$Br$_x$)$_3$ and 10x10x10 for MAPbI$_2$Br and MAPbIBr$_2$. The crystal structure of MAPbBr$_3$ from Ref. \cite{Walsh_collection} was used as a starting point to construct the 2x2x2 supercells and replace the required number of Br atoms by I. The geometry optimization was performed using the PBE functional because it has been shown \cite{Lindmaa} that AK13 is not accurate for this kind of optimization. The convergence thresholds on the total energy and forces for the ionic minimization were set to 1.0x10$^{-4}$ and 1.0x10$^{-3}$ a.u., respectively. The Broyden-Fletcher-Goldfarb-Shanno quasi-newton algorithm was used for cell dynamics.

\section{Conclusions}

The employment of the AK13/GAM combination in the DFT calculations of the electronic structure of the MAPb(I$_{1-x}$Br$_x$)$_3$ series allowed for an accurate estimation of the band gaps as compared with reported experimental data. Calculated DOS reveals differences in the distribution of the I and Br $p$ states as dominating components in the valence band of MAPb(I$_{1-x}$Br$_x$)$_3$. The gravity center of the Br $p$ DOS appears to be at lower energies and the I $p$ states tend to contribute more in vicinity of VBM, at least for $x\leq0.5$. Such a behavior was found to be in agreement with observed changes in the shape of the XPS spectra of the valence band for this series upon varying the Br content. The calculated band gap size in MAPb(I$_{1-x}$Br$_x$)$_3$ shows a quadratic-like dependence on the $x$ value in the cubic crystal structure region and the derived bowing parameter supports the experimental reports claiming it being significant.

\textbf{Notes}
The author declares no competing financial interest.

\begin{acknowledgement}
The author acknowledges the support from the Swedish Research Council (research grant 2018-05525). The computations and data handling were enabled by resources provided by the Swedish National Infrastructure for Computing (SNIC) at National Supercomputer Centre at Link\"{o}ping University partially funded by the Swedish Research Council through grant agreement no. 2018-05973.
\end{acknowledgement}

\bibliography{MAPbBr3-xIx}

\end{document}